\documentclass{kapproc} 

\normallatexbib

\usepackage{graphicx}
\newcommand{\be}{\begin{equation}}
\newcommand{\ee}{\end{equation}}
\begin{document}



\articletitle[]{Nuclear equation of state and the structure of neutron stars}


\author{A.E.L. Dieperink}
\affil {Kernfysisch Versneller Instituut, Zernikelaan 25, NL-9747AA Groningen, The Netherlands}
\author{D. Van Neck and Y. Dewulf}
\affil{Laboratory of Theoretical Physics, Ghent University,
Proeftuinstraat 86, B-9000 Gent, Belgium}
\author{V. Rodin}
\affil{Institut f\"ur Theoretische Physik der Universit\"at T\"ubingen,
D-72076  T\"ubingen, Germany}


 \begin{abstract}
 The hadronic equation of state for a neutron star is discussed
with a particular emphasis  on the symmetry energy. The results of several
microscopic approaches are compared and also a new calculation in terms
of the self-consistent Green function method is presented. In addition
possible constraints on the symmetry energy coming from empirical information
on  the neutron skin of finite nuclei are considered.
 \end{abstract}

\section{Introduction}
In describing properties of neutron stars the equation of state
(EoS) plays a crucial role as an input.
 During this workshop the effects of
QCD degrees of freedom which become important at higher densities
have been discussed in great detail. However, a quantitative
understanding of the EoS also at lower densities is a prerequisite for
the description of neutron stars properties. Conventional nuclear
physics methods (i.e., in terms of hadronic degrees of freedom)
are still being improved both conceptually as well as numerically,
but all microscopic many-body approaches do require some sort of
approximation scheme.

The aim of this contribution is to review the present understanding
of the hadronic EoS, and the nuclear symmetry energy (SE) in particular.
In the past the latter quantity has been computed mostly in terms of the 
Brueckner-Hartree-Fock (BHF) approach. In order to get an idea about
the accuracy of the BHF result we present a
recent calculation \cite{Die03}
in terms of the self-consistent Green function method, that can be
considered as a generalization of the  BHF approach. The results
are compared with those from various  many-body approaches, such as
variational and relativistic mean field approaches.
In view of the large spread in the theoretical predictions we also
examine possible constraints on the nuclear SE that may be obtained 
from information from finite nuclei (such the neutron skin).

\section{Equation of state and symmetry energy}
The central quantity which determines neutron star properties is
the EoS, at $T=0 $ specified by the energy density
$\epsilon(\rho).$ For densities above, say, $\rho= 0.1$ fm$^{-3}$
one assumes a charge neutral uniform matter consisting of protons,
neutrons, electrons and muons; the conditions imposed are charge
neutrality, $ \rho_p=\rho_e +\rho_\mu, $ and beta equilibrium,
$ \mu_n=\mu_p +\mu_e $ with $\mu_e=\mu_\mu.$   \\
Given  the energy density $\epsilon= \epsilon_N+ \epsilon_e+ \epsilon_\mu $
the total pressure, $P=  P_N+ P_e +P_\mu,$ is obtained 
as $P= P(\epsilon/c^2)= \sum \rho_i\mu_i - \epsilon, $
with the chemical potentials given by
$\mu_i= \frac{\partial \epsilon}{\partial \rho_i} $.

\begin{figure}[ht]
\includegraphics[width=10cm]{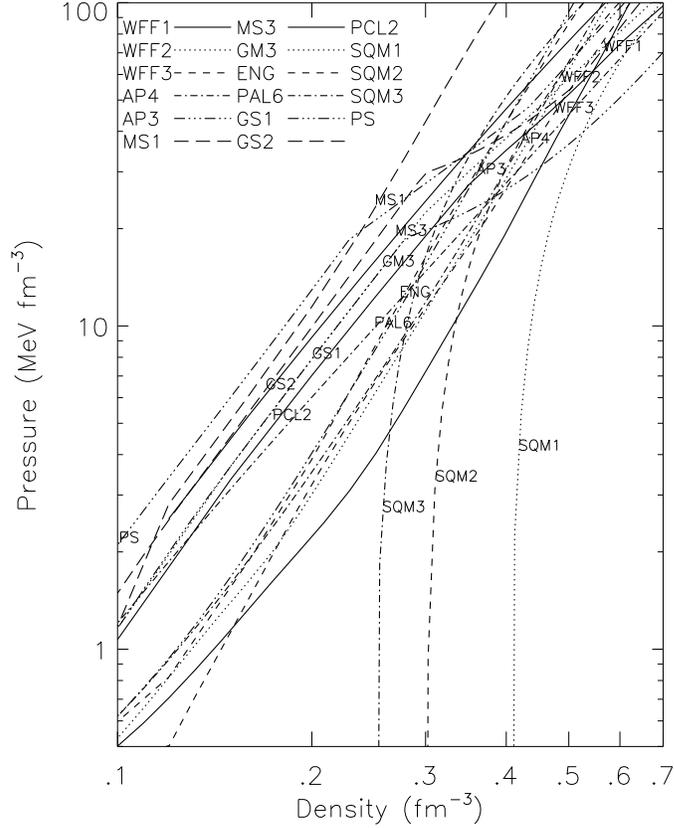}
\caption{Pressure versus density for a variety of EoS; taken from
\cite{Latt} where explanation of symbols is given.}    \label{Latt-eos}
\end{figure}

To illustrate the present status of the  
situation in Fig.~\ref{Latt-eos}  a compilation made by the Stony
Brook group \cite{Latt} of the relation between the pressure and
density for a wide variety of EoSs (but still a tiny fraction of
all available calculations) is shown.
This figure shows that there is an appreciable spreading in the
calculations of the pressure. Qualitatively one can distinguish  three
different classes. First the  EoSs that correspond to
self-bound systems (such as ``strange quark matter") have the property
that the pressure does not vanish at zero
density.   Then there are two globally  parallel  bands.
It appears that most conventional non-relativistic approaches fall in 
the lower
(softer) band and the covariant results in the upper (more rigid)
band. It is worth noting  that even at densities  around and below 
saturation, $\rho_0 \sim
0.17$ fm$^{-3},$ the predictions vary appreciably (about a factor
4). At higher densities some models show a sudden change in the
slope of the pressure. This behavior can be ascribed to the onset of new
degrees of freedom, e.g., the appearance of hyperons, kaons and/or
quarks which lead to a softening.

 It is well known that at lower densities the properties of the EoS  are 
primarily determined by
 the SE \cite{Latt}. 
 The latter is defined in terms  of a Taylor series  expansion of the
energy per particle for nuclear matter in terms of the asymmetry
parameter $\alpha=(N-Z)/A $ (or equivalently the proton fraction
$x=Z/A$),
 \be E(\rho,\alpha)=
E(\rho,0) +S_2(\rho)\alpha^2 + S_4(\rho)\alpha^4+ \ldots,
\label{BE} \ee  where $S_2$ is the quadratic SE. 
It has been shown \cite{Zuo,Zuo02} that the deviation
from the parabolic law in Eq.~(\ref{BE}), i.e., the term corresponding to 
$S_4,$ is quite small.

Near the saturation density $\rho_0$ 
 the quadratic SE is expanded as
 \be S_2(\rho)= \frac{1}{2}
\frac{\partial^2E(\rho,\alpha)}{\partial \alpha^2}|_{\alpha=0} =
a_4+ \frac{p_0}{\rho_0^2} (\rho-\rho_0) + \frac{\Delta
K}{18\rho_0^2}(\rho-\rho_0)^2+\ldots  \label{esymm}. \ee The
quantity $a_4$ corresponds to the SE at equilibrium density and
the slope parameter $p_0$ governs the density dependence.

As a result the pressure can be written  as
 \be P(\rho,x)=
\rho^2 \frac{\partial E(\rho,x)} {\partial \rho} = \rho^2 [
E'(\rho, 1/2) +S_2'(\rho) (1-2x)^2 +\ldots ]. \label{press} \ee By
using beta equilibrium in a neutron star, $\mu_e= \mu_n-\mu_p=
-\frac{\partial E(\rho,x)}{\partial x} \sim  S_2(\rho) (1- 2x) ,
$ and the result for the
 electron chemical potential, $\mu_e=3/4\hbar cx(3\pi^2\rho x)^{1/3}, $ one
finds the proton fraction at saturation density,
to be quite small,
 $x_0 \sim 0.04.$
Hence the  pressure at saturation density  can in good  approximation be 
expressed in terms of (the density dependence of) the SE
\be P(\rho_0) = \rho_0(1-2x_0)(\rho_0S_2'(\rho_0)(1-2x_0)+ S_2(\rho_0)
x_0) \sim \rho^2_0 S_2'(\rho_0). \ee


\section{How well do we know the symmetry energy?}
 In the following  the present status of
 calculations of the SE is reviewed; first the ones in which a microscopic
 nucleon-nucleon (NN) interaction is used, followed by the 
phenomenological mean field approaches. In section 4 we
 discuss possible constraints that can be obtained from empirical
information.


\subsection{The SE in the BHF scheme}
In the Brueckner-Hartree-Fock (BHF) approximation, the
Brueckner-Bethe-Goldstone (BBG) hole-line expansion is truncated
at the two-hole-line level \cite{BB01}. The short-range NN repulsion is
treated by a resummation of the particle-particle ladder diagrams
into an effective interaction or $G$-matrix. Self-consistency is
required at the level of the BHF single-particle spectrum
$\epsilon^{BHF}(k)$,
\begin{equation}
\epsilon^{BHF}(k)=\frac{k^2}{2m}+\sum_{k'<k_F}\mbox{Re}
<kk'|G(\omega=\epsilon^{BHF}(k)+\epsilon^{BHF}(k'))|kk'> .
\label{eq:spectrum}
\end{equation}
In the standard choice BHF the self-consistency requirement
(\ref{eq:spectrum}) is restricted to hole states ($k<k_F,$ the Fermi 
momentum) only,
while the free spectrum is kept for particle states $k>k_F$. The
resulting gap in the s.p.\ spectrum at $k=k_F$ is avoided in the
continuous-choice BHF (ccBHF), where Eq.~(\ref{eq:spectrum}) is
used for both hole and particle states. The continuous choice for
the s.p.\ spectrum is closer in spirit to many-body Green's
function perturbation theory (see below). Moreover, recent results
indicate \cite{Bal00,Bal01} that the contribution of higher-order
terms in the hole-line expansion is considerably smaller if the
continuous choice is used.

Although the BHF approach has several shortcomings  it provides  a numerically simple and convenient
scheme to provide insight in some aspects of the symmetry energy.

\subsubsection{Decomposition of SE}
Some insight into the microscopic origin of the SE
can be obtained by examining the
separate contributions to the  kinetic and potential energy \cite{Zuo}.

\begin{figure}[ht]
\includegraphics[width=0.6\textwidth]{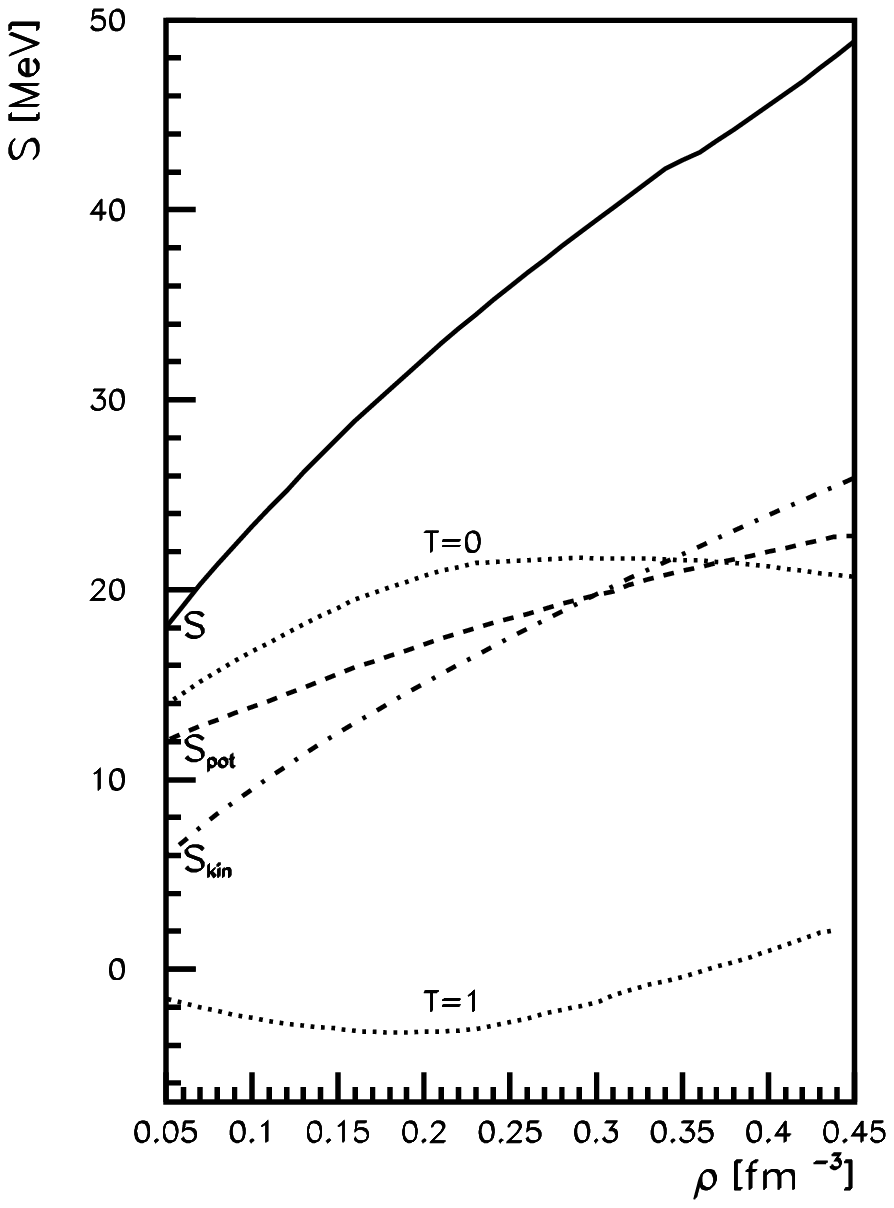}
\vspace*{-6cm}
 \narrowcaption{The SE $S_2$ (full line) and the
contributions to $S_2$ from the kinetic (dash-dotted line) and
potential energy (dashed line), calculated  within a ccBHF scheme
and using the Reid93 interaction. Also shown (dotted lines) are
the $T=0$ and $T=1$ components of the potential energy
contribution. \label{fig:bhf}}
\end{figure}

The results of a ccBHF calculation \cite{Die03} with the Reid93
interaction, including partial waves with $J<4$ in the calculation
of the G-matrix are presented in Fig.~\ref{fig:bhf}, where the SE
is decomposed into various contributions
shown as a function of the density $\rho$.
The kinetic energy contribution, $S_{\rm kin},$ to the SE in BHF
  is given by the free Fermi-gas expression
(it differs from the standard one, which is
based upon
 the derivative rather than  the finite difference)
\begin{equation}
S_{\rm kin}=E_{\rm kin,PNM}-E_{\rm kin,SNM}=
\frac{3}{10m}(3\pi^2)^{\frac{2}{3}}\rho^{\frac{2}{3}}
\left(1-{2^{-\frac{2}{3}}}\right),
\end{equation}
and it determines to a large extent the density dependence of the SE.
In Fig.~\ref{fig:bhf} we also show the symmetry potential $S_{\rm pot}
= S_2 -S_{\rm kin}$, which is much flatter, and the contributions to
$S_{\rm pot}$ from both the isoscalar ($T=0$) and isovector ($T=1$)
components of the interaction. Over the considered density range
$S_{\rm pot}$ is dominated by the positive $T=0$ part. The $T=0$
partial waves, containing the tensor force in the  $^3S_1$-$^3D_1$
channel which gives a major contribution to the potential energy
in SNM, do not contribute to the PNM energy. The $T=0$
contribution peaks at $\rho\approx 0.3$ fm$^{-3}$. The decrease of
this contribution at higher densities is compensated by the
increase of the $T=1$ potential energy, with as a net result a
much weaker density-dependence of the total potential energy.


\subsubsection{Dependence of the SE on NN interaction}
Engvik et al.\ \cite{Engvik} have performed lowest-order
BHF
calculations in SNM and PNM for all
``modern'' potentials (CD-Bonn, Argonne v18, Reid93, N{ij}megen I
and II), which fit the N{ij}megen NN scattering database with high
accuracy. They concluded that for small and normal densities the
SE is largely independent of the interaction used, e.g.\ at
$\rho_0$ the value of $a_4$ varies around an average value of
$a_4$=29.8 MeV by about 1 MeV. At larger densities the spread
becomes larger; however, the SE keeps increasing with density, in
contrast to some of the older potentials like Argonne v14 and the
original Reid interaction (Reid68) for which $S(\rho )$ tends to
saturate at densities larger than $\rho =0.4$ fm$^{-3}$.


\subsection{Variational approach}
 Detailed studies for SNM and PNM using variational
chain summation (VCS) techniques were performed by Wiringa et al.\
\cite{Wir} for the Argonne Av14 
NN interaction in
combination with the Urbana UVIII three-nucleon interaction (3NI),
and by Akmal et al.\ \cite{Akmal} for the modern Av18 NN potential
in combination with the UIX-3NI.

It should be noted  that the results of VCS and BHF calculations
using the same NN interaction  disagree in several aspects. For instance 
for SNM the  VCS and BHF calculations 
saturate at different values of density \cite{BB01}. As for the
SE using only two-body interactions  the VCS approach  
 yields a smaller value for $a_4$ than  BHF (see Table~\ref{Effect3NI}),
 and  as a function of density
 in VCS the SE levels off at $\rho \sim 0.6$ fm$^{-3},$ whereas
 the BHF result continues to increase.
 Therefore it seems natural to ask  whether the inclusion of more
correlations by  extending the BHF method (which is basically a mean field 
approximation) will lead to results closer to those of  VCS.                                                  

\subsection{Self-consistent Green function method}
 As noted above BHF  has several deficiencies: it does not saturate at
the empirical density, and it violates the Hugenholz-van Hove theorem.

In recent years several groups have considered the replacement of
the BBG hole-line expansion with self-consistent Green's function
(SCGF) theory \cite{Bozek,Dewulf01,Dewulf02}. In
ref.\cite{Dewulf02,Dewulf03} the binding energy for SNM was calculated
within the SCGF framework  using the Reid93 potential. In
ref.\cite{Die03} we have extended these calculations to PNM and
considered the corresponding SE.

The SCGF approach differs in two important ways from the BHF
scheme. Firstly, within SCGF particles and holes are treated on an
equal footing, whereas in BHF only intermediate particle $(k>k_F)$
states are included in the ladder diagrams. This aspect ensures
thermodynamic consistency, e.g.\ the Fermi energy or chemical
potential of the nucleons  equals the binding energy at saturation
(i.e. it fulfills the Hugenholz-van Hove theorem). In the
low-density limit BHF and SCGF coincide. As the density increases
the phase space for hole-hole propagation is no longer negligible,
and this leads to an important repulsive effect on the total
energy. Second, the SCGF generates realistic spectral functions,
which are used to evaluate the effective interaction and the
corresponding nucleon self-energy. The spectral functions exhibits
a depletion of the quasi-particle peak and the appearance of
single-particle strength at large values of energy and momentum,
in agreement with experimental information from $(e,e'p)$
reactions. This is in contrast with the BHF approach where all
s.p.\ strength is concentrated at the BHF-energy as determined
from Eq. (\ref{eq:spectrum}).

In the SCGF approach the particle states ($k>k_F$), which are
 absent in the BHF energy sum rule
do contribute according to the energy sum rule ($d=1(2)$ for
PNM(SNM))
\begin{equation}
\frac{E}{A} = \frac{d}{\rho} \int \frac{d^3k}{(2\pi)^3}
\int_{-\infty}^{\varepsilon_F} d\omega\ \left( \frac{k^2}{2m} +
\omega \right) S_h(k,\omega) , \label{eq:epps}
\end{equation}
expressed in terms of the nucleon spectral function $S_h
(k,\omega)$.

\begin{figure}[ht]
\centering{
\includegraphics[width=0.95\textwidth,height=0.5\textwidth] {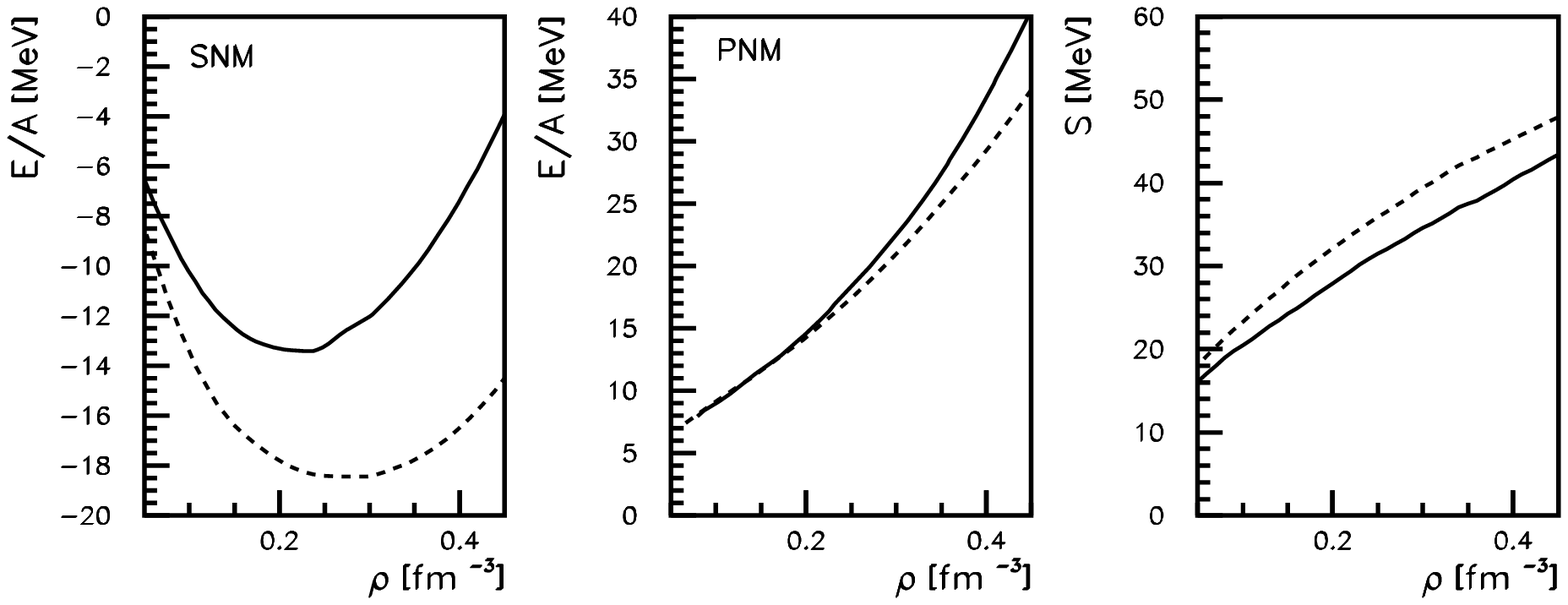}
\caption{The energy per particle for symmetric nuclear matter
(left panel) and pure neutron matter (central panel) for the
Reid93 interaction. The dashed line refers to a ccBHF calculation,
the full line to a SCGF calculation. The right panel displays the
SE in these two approaches.} \label{fig:symbhf}}
\end{figure}

To illustrate the difference between the ccBHF and SCGF approaches
the results for both SNM and PNM are compared in the left and
central panels of Fig.~\ref{fig:symbhf} for the Reid93 interaction.
One sees that the inclusion of high-momentum nucleons leads
roughly to a doubling of the kinetic and potential energy in SNM,
as compared to BHF. The net effect for the total energy of SNM is
a repulsion, increasing with density \cite{Dewulf02}. This leads
to a stiffer equation of state, and a shift of the SNM saturation
density towards lower densities. The above effects which are
dominated by the tensor force (the isoscalar $^3S_1$-$^3D_1$ partial
wave)  are  much smaller in PNM.

The corresponding SE, shown in the right panel of
Fig.~\ref{fig:symbhf}, is dominated by the shift in the total
energy for SNM, and lies below the ccBHF symmetry energy in the
entire density-range. At $\rho_0 =0.16 $ fm$^{-3}$ the parameter
$a_4$ is reduced from 28.9 MeV to 24.9 MeV, while the slope $p_0$
remains almost the same (2.0 MeVfm$^{-3}$ compared to 1.9
MeVfm$^{-3}$ in BHF).

\subsection{Three-body force in microscopic approaches}
As noted above at higher densities the EoS is sensitive to 3NF contributions.
Whereas the 3NF for low densities seems now well understood its
contribution to nuclear matter densities remains  unsettled. 
In practice in calculations of the symmetry energy in the BHF
approach two types of 3NF have been used in calculations; in
ref.\cite{Zuo02} the microscopic 3NF based upon meson exchange by
Grang\'e et al. was used, and in ref.\cite{Baldo97} as well in most
VCS calculations the Urbana
interaction. The latter has in addition to an attractive
microscopic two-pion exchange part a repulsive phenomenological
part constructed in such a way that the empirical saturation point
for SNM is reproduced.
Also in practice in the BHF approach  to simplify the computational efforts
 the  3NF 
is reduced to a density dependent two-body force  by averaging over
the position of the third particle.

In general  inclusion of the 3NF stiffens the SE for $\rho > \rho_0.$
For example, with the Av18+UIX interaction  $a_4$  increases from 27 to 30 MeV
and $p_0$ from 2 to 3 MeVfm$^{-3}$
(see Table \ref{Effect3NI}).

In view of the important contribution from the 3NF to the SE at higher densities 
there seems a clear need to get a more quantitative understanding of the
repulsive part of the 3NF in PNM.
\subsection{Dirac-Brueckner-Hartree-Fock}
The SE has also been computed in the Dirac-Brueckner-Hartree-Fock
(DBHF) approach \cite{Lee, Toki}. 
Although  some of the shortcomings of  BHF
(mean field treatment of the Pauli operator and violation
of Hugenholz-van Hove theorem) persist in the DBHF method
it appears that in the latter approach one is able to
 reproduce  the  saturation properties of SNM without the inclusion of
a three-body interaction. 
A general feature of relativistic methods is that the
SE   increases almost linearly with density,
and more rapidly than in the non-relativistic case. This
difference can be attributed to two effects. First the covariant
kinetic energy which is inversely proportional to $\sqrt{k_F^2+
m^{*2}}$  is larger because of the decreasing Dirac mass, $m^*$,
with increasing density. Secondly the contribution from
rho-exchange appears to be larger than in the non-relativistic
case \cite{Lee}.

\subsection{Relativistic mean-field approach}

Relativistic mean field (RMF) models have been applied succesfully to 
describe properties of finite nuclei.
In general ground state  energies, spin-orbit splittings, etc.  can be 
described well in terms of a few parameters ref.\cite{Ring}.
Recently it has lead to the suggestion that the bulk SE is strongly
correlated with the neutron skin \cite{Brown,Furn}  (see below).
In essence the 
method is based upon the use of energy-density functional (EDF) theory.


In practice covariant approaches are formulated  either 
in terms of a  covariant
lagrangian with $\sigma,\ \omega$ and $\rho$ exchange (and
possibly other mesons) \cite{Ring, Horo02}, or in terms of
contact interactions \cite{Furn}, solved as an EDF in the 
Hartree-Fock approximation. Sets of  model
parameters are determined by fitting bound state properties of
nuclei. Specifically the isovector degree of freedom is governed
by the exchange of isovector mesons; in case of $\rho$-meson
exchange
 the (positive definite)  contribution to the SE  is given by
 \be a_4= \frac{k^2_F}{6 \sqrt{m^{*2}+k_F^2}}+ \frac{g^2_\rho}{8m^{2}_\rho} \rho_0,
 \ee
and its potential energy contribution to $p_0$, which scales with
that for $a_4$, is
   $ \frac{g^2_\rho}{8m^{2}_\rho}$  \cite{Furn}.
Typical values obtained for $p_0$ are around 4-6 MeV fm$^{-3}$,
and $ a_4\sim $ 30-36 MeV, i.e., considerably larger than in
non-relativistic approaches
 (a large part of the enhancement can
be ascribed to the fact that the kinetic contribution is larger,
because $m^* <m$).

Recently   this approach was extended  by inclusion of  the
isovector-scalar partner, the $\delta$-meson, of the isoscalar
scalar $\sigma$-meson \cite{Liu02}. Unfortunately the value of
the coupling for the $\delta$-meson cannot be determined well by
fitting  properties of stable nuclei.
 Also in its simplest, density independent form, the inclusion of the 
$\delta$-meson
 leads to an even larger net value for $p_0$.
This happens because of
the presence of the Lorentz factor $m^*/E$ in the scalar potential
contribution, $\sim -\frac{g_\delta^2}{8m_\delta^2} \frac{m^*}{E},
$ which decreases with increasing density.

\subsection{Effective field theory}
Recently the density dependence of the symmetry energy has been
computed in chiral perturbation effective field theory, described
by   pions plus one cutoff parameter, $\Lambda,$ to simulate the
short distance behavior \cite{Kais}. The nuclear matter
calculations have been performed up to three-loop order; the
density dependence comes from the replacement of the free nucleon propagator
by the in-medium one, specified by the Fermi momentum $k_F$
$$(\slash \hspace{-2.2mm} p +M)
\left( \frac{i}{p^2-M^2+i\epsilon}-2\pi\delta(p^2-M^2)
\theta(p_0)\theta(k_F-p)\right). $$
The resulting EoS is expressed as an expansion in powers of  $k_F$ and
the value of $\Lambda \approx 0.65$ GeV is adjusted to the
empirical binding energy per nucleon. 
In its present form the validity of this
approach is clearly confined to relatively small values of the
Fermi momentum, i.e. rather low densities.
Remarkably for SNM the calculation appears to be able to reproduce the
microscopic EoS up to $\rho \sim 0.5$ fm$^{-3}$.
As for the SE the value obtained in this
approach for $a_4= 33$ MeV is in reasonable agreement with the
empirical one; 
however, at higher densities ($\rho > 0.2$ fm$^{-3}$)  a downward bending
is predicted (see Fig.~\ref{Overview-Srho}) which is not present in other 
approaches.

\subsection{Comparison of results}

\begin{figure}[ht]
\centering{
\includegraphics[width=0.7\textwidth,height=0.7\textwidth] {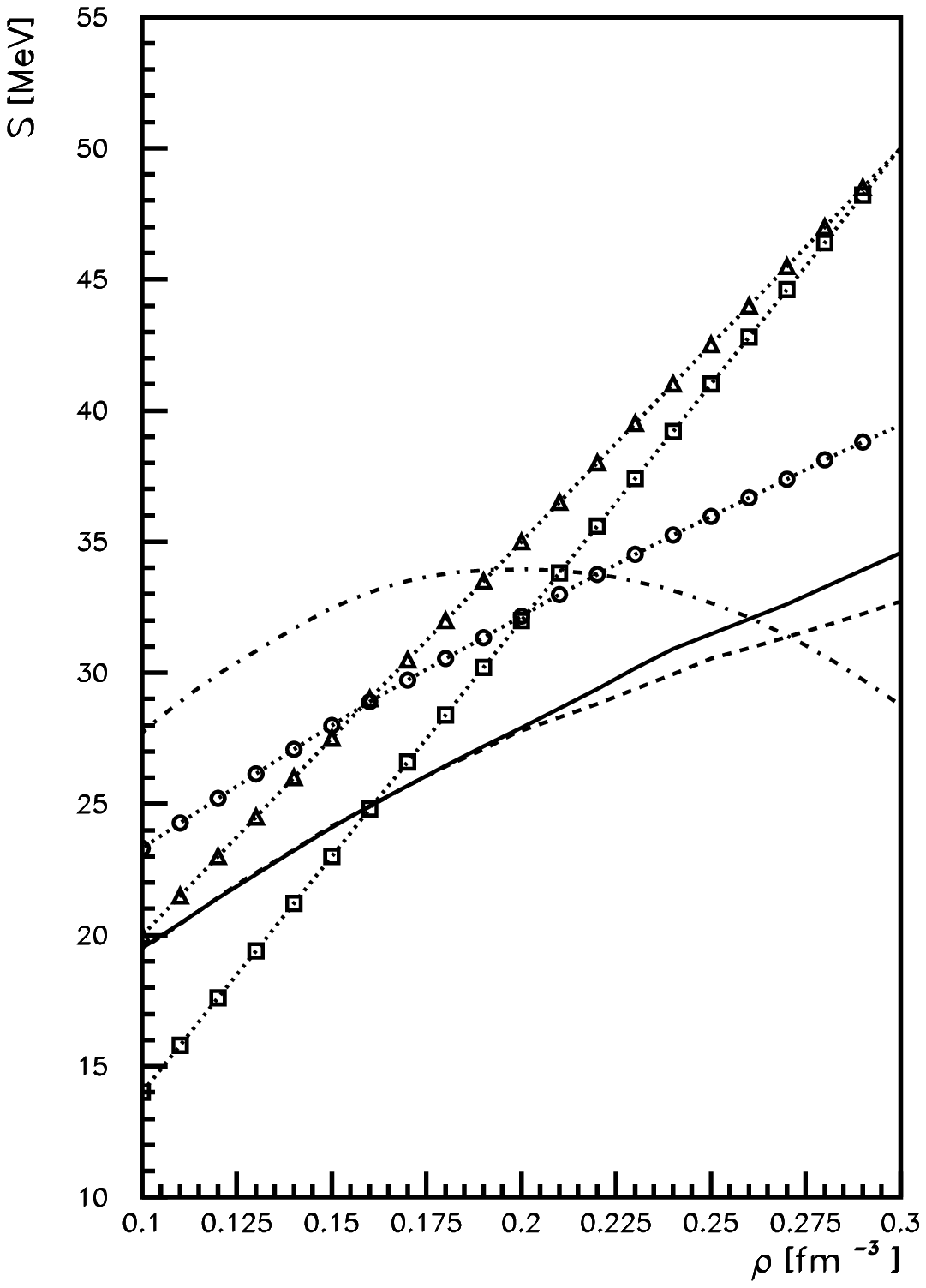}
\caption{ Overview of several theoretical predictions for the
SE: Brueckner-Hartree-Fock (continuous choice)
with Reid93 potential (circles), self-consistent Green function
theory with Reid93 potential (full line), variational calculation
from \cite{Wir} with Argonne Av14 potential (dashed line),
DBHF calculation from \cite{Lee}
(triangles), relativistic mean-field model from \cite{Liu02}
(squares), effective field theory from \cite{Kais} (dash-dotted
line).} }
\label{Overview-Srho}
\end{figure}

To summarize the present situation in Fig.~\ref{Overview-Srho}
the resulting density dependence of the SE for the approaches  discussed 
above
are compared (excluding the 3NF contribution). One sees that the 
covariant models
predict a much larger increase of the SE with the density than the
non-relativistic approaches. The lowest-order BHF method predicts  a 
somewhat higher 
value for $a_4$  than both the VCS and  SCGF methods,
which lead to very similar results; whether that can be ascribed
to a consistent treatment of correlations in these methods, or is 
fortuitous, is not clear.   

\begin{table} 
\caption{Results for the symmetry energy parameters $a_4$ (in MeV) 
and  $p_0$ (in MeVfm$^{-3}$)}
\begin{center}
\begin{tabular}{c|c|c|c|c|c|c|c|}
\hline
 & BHF & BHF+3NF\cite{BB01} & VCS\cite{Akmal} & VCS+3NF\cite{Akmal} &SCGF\cite{Die03} & 
DBHF\cite{Lee} & RMF\cite{Furn} \\
\hline
$a_4$ & 28.9 &  30  &27 & 30   &  24.9 &31  & 30-36 \\
$p_0$  & 1.9 &  2.9 &2.0 &3.1  & 2.0  & 3.0 & 4-6  \\
\hline
\end{tabular}
\end{center}
\label{Effect3NI}
\end{table}
Clearly at least part of the differences should be attributed to the 
different
selection of the  mesonic degrees of freedom  in the various 
models.
In the microscopic  approaches  the tensor force
mediated by $\pi$ and $\rho$ exchange seems to play the dominant role. 
On the other hand in the mean field approach
   explicit pion exchange is usually not included and hence
there the isovector effect solely comes from the shorter range 
$\rho$-exchange.
In fact it has  been  argued that in contrast
to isoscalar properties the long-range pion exchange should play an
essential role in determining the isovector properties \cite{Furn}.

\section{Empirical information on the SE}
In view of the existing uncertainties in the
calculation of the SE one may ask whether from finite nuclei one can 
obtain  
experimental constraints on the symmetry energy as a function of
density. In this section some recent activities
pertaining to this issue are reviewed.
\subsection{Relation between SE and neutron skin} 
\label{Relationship}
Recently 
 in applying the non-relativistic Skyrme  Hartree-Fock
(SHF) model Brown  \cite{Brown}  noted that  certain combinations 
of parameters in the SHF are not well determined  by a fit to ground
state binding energies alone; as a result  a wide range of predictions 
for the  EoS for PNM can be obtained. At the same time he found a
correlation between the derivative of the neutron star EoS (i.e.,
basically the symmetry pressure $p_0$) and the neutron
skin in $^{208}$Pb.

\begin{figure}[t]
 \vspace*{-0.6cm}
 \begin{center}
 \includegraphics[width=8cm,angle=270]{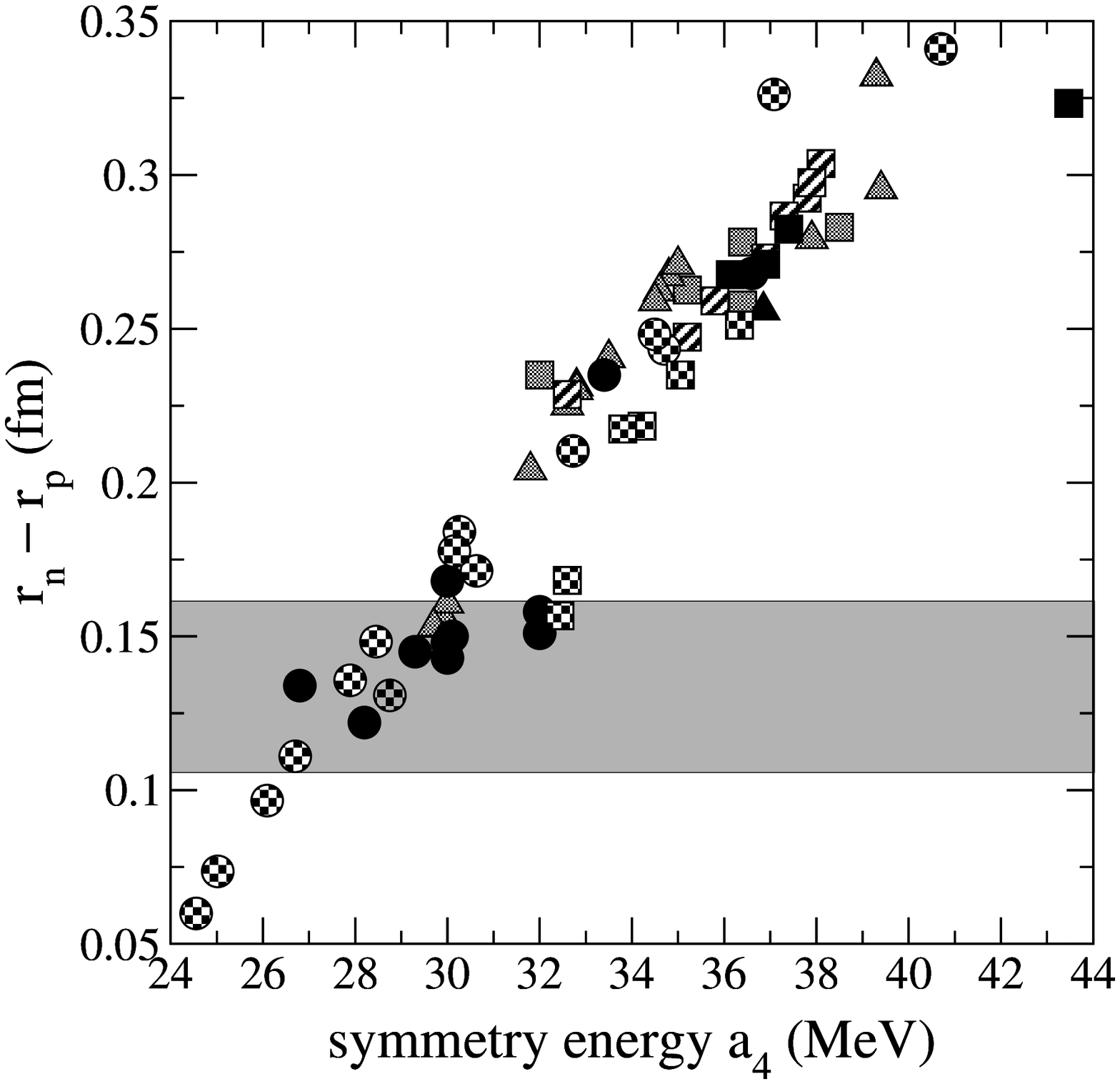}
 \caption{Neutron skin thickness versus $a_4$ for $^{208}$Pb for a variety of
mean field models (from 
\cite{Furn}).
 The circles correspond to results for the Skyrme force, the squares to RMF models
 with mesons,
and  the triangles to RMF with point couplings;
  the shaded area  indicates the range of $\Delta R$ values  consistent with the present 
empirical information for $^{208}$Pb.}

     \label{DeltaR-vs-a4}
 \end{center}
 \end{figure}

 Subsequently Furnstahl \cite{Furn} in a more extensive study
 pointed out 
that within the
framework of mean field models
(both non-relativistic Skyrme as well as relativistic models)
 there exists an
almost linear  empirical correlation between theoretical
predictions for both $a_4$ and its density dependence, $p_0, $ and
the neutron skin, $\Delta R=R_n-R_p,$ in heavy nuclei.
 This is illustrated for $^{208}$Pb in Fig.~\ref{DeltaR-vs-a4} (from 
ref.\cite{Furn};
 a similar correlation is found between $\Delta R $ and $p_0$).
Note that whereas the Skyrme results cover a wide range of $\Delta R$ values the
RMF predictions in general lead to $\Delta R > 0.20$ fm. 

The interpretation of the nuclear matter results in part depends on the
question whether there is a surface  contribution to the SE
in  finite nuclei. In ref.\cite{Oya} it  was found that for heavy nuclei 
the
latter is of minor importance, which has  also been confirmed in
ref.\cite{Furn}.

\subsection{ Insight in the correlation of $\Delta R$ and $a_4$}
 The above observation suggests an intriguing
relationship between a bulk property of infinite nuclear matter
and a surface property of finite systems.
\\ Here we want to point out that this correlation can be understood
naturally in terms of the Landau-Migdal approach. To this end we consider a
simple mean-field model (see, e.g., ref.\cite{Gor00}) with the
Hamiltonian consisting of the single-particle mean field part
$\hat H_0$ and the residual particle-hole interaction $\hat
H_{ph}$:
\be
\hat H= \sum\limits_a (T_a+U_0(x_a)+U_1(x_a)+U_C(x_a)) +\hat H_{ph}, \label{1.1} \ee
where  $(x=(r,\sigma,\tau))$
\be U_0(x)=U_0(r)+U_{so}(x);\ U_1(x)=\frac12 S_{\rm
pot}(r)\tau^{(3)};\ U_C(x)=\frac12 U_C(r)(1-\tau^{(3)}).
 \nonumber
\ee
Here, the mean field potential 
includes the
phenomenological isoscalar part $U_0(x)$ along with the isovector
$U_1(x)$ and the Coulomb $U_C(x)$ parts calculated consistently in
the Hartree approximation; $U_0(r)$ and
$U_{so}(x)=U_{so}(r)\vec{\sigma} \cdot\vec l$ are the central and
spin-orbit parts of the isoscalar mean field, respectively, and
$S_{\rm {pot}}(r)$ is
 the potential part of the symmetry energy.
\\ In the Landau-Migdal approach the effective isovector particle-hole
interaction $\hat H_{ph}$ is given by
\begin{equation}
\hat H_{ph}=\sum\limits_{a>b}
(F'+G'\vec\sigma_a\vec\sigma_b)\vec\tau_a\vec\tau_b
\delta(\vec{r}_a-\vec{r}_b), \label{1.3}
\end{equation}
where $F'$ and $G'$ are the phenomenological Landau-Migdal
parameters.

The model Hamiltonian $\hat H$ in Eq.(\ref{1.1}) preserves isospin
symmetry if the condition
 \be
\label{1.4} [\hat H, \hat   T^{(-)}]=\hat U^{(-)}_C,
\label{commHT} \ee is fulfilled, where
$ \hat T^{(-)}=\sum\limits_a \tau_a^{(-)},\ \hat
U^{(-)}_C=\sum\limits_a U_C(r_a)\tau_a^{(-)}. $
  With the use of  Eqs.~(\ref{1.1}),(\ref{1.3}) 
    the condition eq.(\ref{commHT})  in the random phase approximation (RPA)
 formalism  leads to
a self-consistency relation between  the   symmetry potential and
the Landau parameter $F'$ \cite{Bir74}:
\begin{eqnarray}
&S_{\rm{pot}}(r)=2F'n^{(-)}(r), \label{1.5}
\end{eqnarray}
where $n^{(-)}(r)=n^{n}(r)-n^{p}(r)$ is the neutron excess
density. Thus, in this model the depth of the symmetry potential
is controlled by the Landau-Migdal parameter $F'$ (analogously to
the role played by the parameter $g_\rho^2$ in relativistic mean
field models).

\begin{figure}[ht]
\includegraphics[width=10cm]{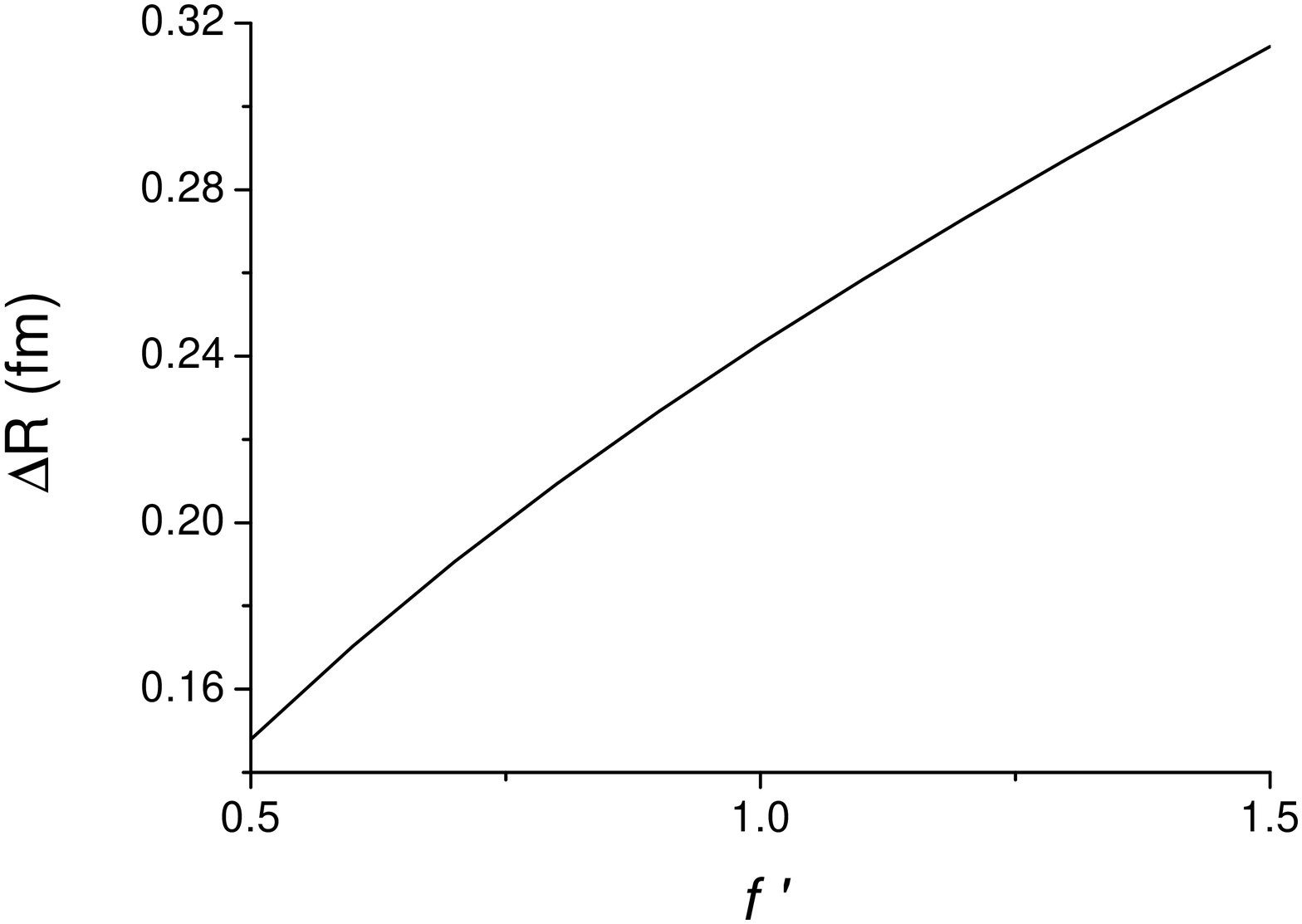}
\caption{Neutron skin in $^{208}$Pb vs. the Landau-Migdal
parameter $f'$.}
 \label{DRvsF'}
\end{figure}

  $S_{\rm{pot}}(r)$ is obtained from Eq.~(\ref{1.5}) by an iterative
procedure;
  the resulting  dependence of $\Delta R$ on the dimensionless
parameter $f'=F'/ (300$ MeV\,fm$^3$)  shown in fig.~\ref{DRvsF'}
  indeed illustrates that
$\Delta R$ depends almost linearly on $f'.$
  Then with the use of the Migdal relation \cite{Migd} which relates the SE
  and $f',$
   \be a_4= \frac{\epsilon_F}{3}(1+2f'), \ee
 an almost linear correlation
between   the symmetry energy, $a_4,$ and the neutron skin is found.

 To get more insight in the role of $f'$ 
 we consider small variations $\delta F'$.
  Neglecting  the variation
of $n^{(-)}(r)$ with respect to $\delta F'$ one has a linear
variation of the symmetry potential: $\delta S_{pot}(r)=2\delta
F'n^{(-)}(r)$. Then in first order perturbation theory, such a
variation of $S_{pot}$ causes the following variation of the
ground-state wave function
\begin{equation}
|\delta 0\rangle=\delta F'\sum\limits_{s} \frac{\langle s | \hat
N^{(-)} | 0\rangle}{E_0-E_s} |s\rangle, \label{1.33}
\end{equation}
with $s$ labeling the eigenstates of the nuclear Hamiltonian and a
single-particle operator $\hat N^{(-)}=\sum\limits_a
n^{(-)}(r_a)\tau^{(3)} _a$. Consequently the variation of the
expectation value of the single-particle operator $\hat
V^{(-)}=\sum\limits_a r^2_a\tau^{(3)}_a$ with $\langle 0 | \hat
V^{(-)} | 0 \rangle=N R^2_n-Z R^2_p$
 can be written as
\begin{equation}
R_p \delta( \Delta R) =\delta F'\frac{2}{A}\sum\limits_{s}
\frac{{\rm Re}\langle 0 | \hat N^{(-)} | s\rangle \langle s | \hat
V^{(-)} | 0\rangle} {E_0-E_s}. \label{deltar2np}
\end{equation}
In practice the sum in Eq.~(\ref{deltar2np}) is exhausted mainly
by the isovector monopole resonance of which the high excitation energy
(about 24 MeV in $^{208}$Pb) justifies the perturbative
consideration. We checked that Eq.~(\ref{deltar2np}) is able to
reproduce directly calculated $\delta (\Delta R)$ shown in
fig.~\ref{DRvsF'} with the accuracy of about 10\%.
\\ As for the correlation between $\Delta R$ and $p_0$ one would need
information on the density dependence of $F'.$ Sofar as we
know it has not been extracted from data on stable nuclei.
In the approximation that $F'$ is density independent one
naturally finds the $p_0$ is proportional to $a_4.$

\subsection{Experimental information on  $\Delta R$ for $^{208}$Pb }
\label{Discussion}
 What are the experimental constraints on the neutron skin?
A  variety of experimental approaches have been explored in the
past to obtain information on $\Delta R.$ To a certain extent all
analysis contain a certain model dependence, which is difficult to
estimate quantitatively. It is not our intention to present a full
overview of existing methods for the special case of $^{208}$Pb.
In particular the results obtained in the past from the analysis of elastic scattering of
 protons and
neutrons have varied depending upon specifics of the analysis
employed. At present the most accurate value for $\Delta R$ comes from a recent detailed
analysis of the elastic proton scattering reaction at $E=0.5-1$
GeV \cite{Clark03}, and of neutron and proton scattering at
$E=40-100$ MeV \cite{Karat02}. For details we refer to these
papers. Here we restrict ourselves to a discussion of the
 some less well known methods that have the potential to provide more accurate
 information on the neutron skin in the future.

\subsubsection{Anti-protonic atoms}
Recently neutron density distributions in a series of nuclei were deduced 
from
anti-protonic atoms \cite{Trzc01}. The basic method determines the
ratio of neutron and proton distributions at large differences by
means of a measurement of the annihilation products which
indicates whether the antiproton was captured on a neutron or a
proton. In the analysis two  assumptions are made. First
 a best fit value for the ratio $R_I$ of the imaginary parts of
the free space $\bar{p}p$ and $\bar{p}n$ scattering lengths equal
to unity is adopted. Secondly in order to reduce the density ratio
at the annihilation side to a a ratio of rms radii a two-parameter
Fermi distribution is assumed. The model dependence introduced by
these assumptions is difficult to judge. Since a large number of
nuclei have been measured one may argue that the value of $R_I$ is
fixed empirically.
 \subsubsection{Parity violating electron scattering}
Recently it has been proposed to use the (parity violating) weak
interaction to probe the neutron distribution. This is probably
the least model dependent approach \cite{Horopv}. The weak
potential between electron and a nucleus \be \tilde{V}(r)=
V(r)+\gamma_5A(r), \ee where the axial potential $A(r)=
\frac{G_F}{2^{3/2}} \rho_W(r). $  The weak charge is mainly
determined by neutrons \be \rho_W(r)= ( 1-4\sin^2\theta_W)
\rho_p(r)- \rho_n(r), \ee with $\sin^2\theta_W  \approx 0.23. $ In
a scattering experiment using polarised electrons one can
determine the cross section asymmetry \cite{Horopv} which comes
from the interference between the $A$ and $V$ contributions. Using
the measured neutron form factor at small finite value of $Q^2 $
and the existing information on the charge distribution one can
uniquely extract the neutron skin. Some slight model dependence
comes from the need to assume a certain radial dependence for the
neutron density,  to extract $R_n$ from a finite $Q^2$ form
factor.
\subsubsection{Giant dipole resonance}
Isovector giant resonances  contain information about the SE
   through the restoring force.
   In particular the excitation of the isovector giant dipole
resonance (GDR) with isoscalar probes has been used to   extract
$\Delta R/R$ \cite{Kras94}. In the distorted wave Born approximation
optical model analysis of
the cross section the neutron and proton transition densities are
needed as an input.
 For example,   in the Goldhaber-Teller picture these are expressed as 
\be g_i(r)= -\kappa \frac{2N_i}{A} \frac{d\rho_i}{dr} \ee 
with $\kappa
$ the oscillation amplitude and $(i=p,n)$; one assumes  ground state
neutron and proton distributions of the form $(x=(N-Z)/A)$ 
\be
\rho_i(r)=\frac{1}{2}(1\pm x \mp \gamma x) \rho( r- c(1 \pm \gamma
x/3)), \ee 
where  $\gamma$ is related to the neutron skin $\Delta R,$ $\gamma = 
\frac{ 3A}{2(N-Z) } \Delta R/ R_0. $
Then for $N>Z$ the
isovector transition density takes on the form $$ \Delta g(r) =
\kappa \gamma \frac{N-Z}{A}( \frac{d\rho}{dr}+\frac{c}{3}
\frac{d^2\rho}{dr^2}), $$ 

 In practice one studies the excitation of the GDR by alpha
particle scattering (isoscalar probe). 
By comparing the experimental   cross section
with the theoretical one  (calculated as a function of the ratio
$\Delta R/ R$) the value of $\Delta R$ can be deduced \cite{Kras94}.

It is difficult to make a quantitative estimate of the uncertainty
in the result coming from the model dependence of the approach. In
the analysis several assumptions must be made, such as the radial
shape of the density oscillations and  the
 actual values of the  optical model parameters.

We note that also other types of isovector giant resonances
have been suggested as a source of information on the
neutron skin, such as the spin-dipole giant resonance \cite{Kras99}
and the isobaric analogue state \cite{Auer72}. At present studies
of these reactions have not led to  quantitative constraints for
the neutron skin of $^{208}$Pb.

\subsubsection{Results for $\Delta R$} 
In  table \ref{summaryDeltaR} we present a
summary of some recent results on $\Delta R$ in $^{208}$Pb. One
sees that  the recent results
are consistent with $\Delta R$ values in the range 0.10-0.16 fm.
It appears from Fig.~\ref{DeltaR-vs-a4} that this range is consistent  with  the conventional
Skyrme model approach but tends to disagree with the results of the RMF
models considered in \cite{Furn}. Also from the correlation plot between
$\Delta R$ and $p_0$ shown in Fig.~11 in ref.\cite{Furn} one may conclude
that a small value for $p_0 \sim 2.0 $ MeVfm$^{-3}$ is favoured over the larger
values from covariant models. 

\begin{table}[ht] 
\begin{center}
\caption{Summary of recent results for $\Delta R$ in $^{208}$Pb}
\begin{tabular}{c|ccc}
method & $\Delta R$ [fm] & error [fm] & ref\\
 \hline
($\vec{p},p^{\prime})$ at 0.5-1.04 GeV   & 0.097 & 0.014 &  \cite{Clark03} \\
nucleon scattering (40-200 MeV) &  0.17 &   & \cite{Karat02} \\
anti-protonic atoms & 0.15 & 0.02 & \cite{Trzc01} \\
 giant dipole resonance excitation  & 0.19 &
0.09 & \cite{Kras94} \\
parity violating electron scattering  & planned & 1\% &
\cite{Horopv}
\\ \hline
\end{tabular}
\end{center}
\label{summaryDeltaR}
\end{table}


\subsection{Information on the SE  from heavy-ion reactions}
 In principle the density dependence of the SE at higher densities (and further away
from $N=Z$)
 can be probed by means of heavy-ion reacties using neutron rich radioactive beams.
 In ref.\cite{Gait03}  possible observable effects from the isovector field
 are considered in terms of the  RMF model.
Of particular interest seems the study of the ratio of $\pi^+/\pi^-$ cross sections
and its energy dependence
which appears sensitive to details of the SE \cite{Li}.
\section{Constraints on EoS from neutron stars}
 Given the EoS as a function of density the mass versus  radius relation 
of a neutron star
 can be obtained in the standard way with the use of the 
Tolman-Oppenheimer-Volkov equation.
Do we need the EoS up to all densities or can we already
draw some conclusions from the lower densities?
In \cite{Latt}
 it was argued that there exists a quantitative  correlation between 
the neutron star  radius and the pressure 
which does not depend strongly on the EoS at the highest densities.
Generally speaking  a stiff (soft)  EoS implies a 
large (small)  radius.

Clearly a simultaneous measurement of the radius and the mass
of the {\it same} star would discriminate between various EoS.
The presently available observations do not yet offer strong constraints. 
For example,  most observed star masses fall in the range of 1.3-1.5 
$M_{\odot}.$ These values seem to rule out the very soft  EoSs predicted
 with hyperons present, which yield $M_{max}\sim 1.3M_{\odot}$ 
\cite{BB01}. This may suggest that either the NY and YY interactions
used as an input must be reexamined, or that the use of the BHF approach 
at high densities is not reliable. 


Other constraints come from   recent observations from X-ray satellites.
Most robust seem  the data from the low mass X-ray binary
EXO 0478-676 obtained by Cottam et al. \cite{Cottam}. From the redshifted
absorption lines from ionized Fe and O a  gravitational redshift $z=0.23$
 was deduced; this gives rise to a mass-to-radius relation
\be M/M_\odot = (1- \frac{1}{ (1+ z)^2}) R/R_{g\odot}    \ee
 with $ R_{g\odot}= 2.95$ km. 
As discussed in detail in various contributions in this workshop
at present  this  constraint
 appears fully  consistent not only with  conventional EoSs, but also 
with most more exotic ones.

\begin{acknowledgments}
This work is part of the research program of the ``Stichting
voor Fundamenteel Onderzoek der Materie" (FOM) with
financial support from the ``Nederlandse Organisatie voor
Wetenschappelijk Onderzoek" (NWO). Y. Dewulf acknowledges
support from FWO-Vlaanderen. 
\end{acknowledgments}







\end{document}